\documentclass[aps,pra,superscriptaddress,twocolumn,10pt]{revtex4-1}
\usepackage{graphicx}
\usepackage{bm}
\usepackage[usenames,dvipsnames,svgnames,table]{xcolor}
\usepackage{epsfig}
\usepackage{amsmath}

\usepackage{amssymb}
\usepackage{array}
\usepackage{nccmath}
\usepackage{dsfont}
\usepackage{array}
\begin{document}
\title{Density matrix of electrons coupled to phonons and the electron-phonon entanglement content of excited states}

\author{Gerg\H o Ro\'osz}
\email{roosz.gergo@wigner.hu}
\affiliation{Wigner Research Centre for Physics, Institute for Solid State Physics and Optics, H-1525 Budapest, P.O. Box 49, Hungary}

\author{Karsten Held}
\email{held@ifp.tuwien.ac.at}
\affiliation{Institute for Solid State Physics,
	 TU Wien, 1040 Vienna, Austria}

\begin{abstract}
  We derive the exact reduced density matrix for the  electrons in
  an analytically solvable  electron-phonon model. Here, the electrons are described as a  Luttinger liquid that is coupled to Einstein phonons.
  %We show that similar singularities   found in [V. M. Stojanovic Phys. Rev. B. 101, 134301 (2020)] by exact diagonalization are present in the thermodynamics limit.  
  We further derive analytical expressions for the electron-phonon entanglement, its spectrum and mutual information  at  finite and zero temperature as well as for excited states. The entanglement entropy is additive in momentum for the quasi-particle excited states, but not in the electron-phonon coupled eigenmodes of the system.
\end{abstract}

\maketitle

\section{Introduction}
The electron-phonon coupling plays an essential role for many phenomena in solid state physics, ranging all the way from  Bardeen–Cooper–Schrieffer(BCS)-type superconductivity \cite{maxwell1950, reynolds1950, allen1950, cooper1956, schrieffer1957, bardeen1957} to  the Peierls instability \cite{peierls1955, lee1973, loss2010} and charge-density-waves~\cite{gruner1988, vanyolos2006, miller2000}. 
Multidimensional  electron-phonon systems are in general not integrable, instead one needs to rely on, for example,  diagrammatic perturbation theory \cite{midgal1958, eliashberg1960}, Monte-Carlo simulations \cite{mishchenko2014, prokofev1998, mishchenko2000} or the tensor-network approach \cite{pino2018}.
In one dimension, the coupled system consisting of a Luttinger liquid and phonons is however integrable \cite{bardeen1951,loss1995,loss2010,wentzel1950}; a series of exact results  \cite{roosz2020,loss1995,loss2010,dora2017, varga1964}
and, in addition, an   accurate variational ansatz exist \cite{weisse2020, ku2002, toyozawa1961, barisic2002}.

In the present work, we derive the reduced density matrix of the electron subsystem in a Luttinger model coupled to Einstein phonons. Previous related work on the density matrix or the entanglement entropy 
includes:  the entanglement between electrons at opposite momenta of the Luttinger liquid \cite{dora2017}, the entanglement entropy  between a single spin and a bosonic bath  \cite{costi2003}, and that
between the electron and the protons in a  $H_2^+$ ion \cite{sanz-vicaro2017}.
For the Holstein model, the entanglement entropy between the electrons and the phonons has been investigated using numerical exact diagonalization  \cite{vladimir2008}, and the thus calculated entanglement spectrum has been used to characterize ground-state nonanalyticities  \cite{vladimir2020}. Also for the Su-Schrieffer-Heeger model \cite{vladimir2008, vladimir2020}  nonanalyticities  in the entanglement spectrum have been found.
In contrast, using the variational method
a continuous cross-over between small and large entanglement in the case of large and small polarons has been reported  \cite{yang2004}.  Further, bosonization has been used to calculate the entanglement entropy, mutual information and entanglement negativity between electrons and acoustic phonons in \cite{roosz2020}.

The entanglement entropy of excited states has become an active area of research in the last years. Among others, in critical spin chains the universal properties of the entanglement entropy has been derived using conformal field theory \cite{berganza2012} and the entanglement content of the Heisenberg chain has been investigated using the Bethe ansatz \cite{molter2014}.
In Ref.~\onlinecite{pizorn2012},
quasi-particle excitations in one-dimensional fermionic system were studied using exact diagonalization and tensor networks, finding  that the entanglement is proportional to the quasi-particle number.
The entanglement of  free fermions in excited states and its  relationship to the  Fermi surface has been investigated in Ref.~\onlinecite{storms2014}, and that of
two-particle excited states in Ref.~\onlinecite{berkovits2013}.   The latter shows that the entanglement entropy is the sum of two terms corresponding to the two particles, if the momentum of the two quasi particles is not too close.
Finally, in Ref.~\onlinecite{alba2009} the entanglement entropy of the excited states of the XY and the Heisenberg XXZ spin chains has been investigated between spatial blocks. 

The above listed works demonstrates, for various special examples with spatial bi-sectioning, that the entanglement  increase by quasi-particle excitations is additive  if the momentum of the quasi-particles is not too close. A unifying derivation in the case of free, homogeneous integrable models of this observation has been performed in a series of works \cite{szecsenyi2018,szechenyi2019_II,szechenyi2019_III}. 
Here, the  entanglement $S$ between two spatial regions  in a state which contains a finite number of quasi-particle excitations with different momenta $q_1 \dots  q_n$  generally is conjectured to be of the form
\begin{eqnarray}
	S= S_{\rm GS} + \sum_{i=1}^n s(q_n) \label{Eq:kdep}
\end{eqnarray} 
where $s(q_n)$ is a model dependent  function which describes the entanglement content of the quasi-particle excitation at momentum $q_n$, and $S_{\rm GS}$ is the ground state entanglement entropy  in the same setting.
Here, we instead investigate a non-spatial bipartition, i.e., the bipartition between the electron and the phonon subsystem and find a similar additivity in momentum space.

The present work is organized as follows: In Section \ref{sec:model} we introduce the model and determine its eigenmodes in Section \ref{sec:cantrans}.
 From these   we derive the reduced density matrix in Section \ref{sec:density_matrix} and present results  on the mutual information and the entanglement entropy at zero and finite temperature in Section \ref{sec:ent_mes} and on excited states  in Section \ref{sec:ent_exc}. Finally, in Section \ref{sec:conc}, we make some closing remarks.

 \section{Luttinger liquid description}
 \label{sec:model}
 In a preceding paper \cite{roosz2020}, we investigated the entanglement between a Luttinger liquid coupled to acoustic phonons. We considered several physically interesting versions of the problem, however not the model corresponding to the Luttinger liquid coupled to Einstein phonons which is at the focus of the present paper.
%\kh{[I did not understand the purpose of the following:]
% ``Therefore we going to summarize here the Luttinger description and the calculation of the R\'{e}nyi-2 entropy of the reduced density matrix here.  The Luttinger description is included in \cite{assaad2008} and we use the same notations in the Hamiltonian.'' [Omit or move somewhere else?]} 
To motivate our model, let us start with the Holstein Hamiltonian  in momentum space which reads
 \begin{align}
 	H&= \sum_q \epsilon(q) c_q^{\dagger} c_q^{{\phantom{\dagger}}} + \omega_0 \sum_q a_q^{\dagger} a_q^{{\phantom{\dagger}}} \nonumber
 	\\
 	&+ \frac{g}{\sqrt{2 \omega_0 M}} \frac{1}{\sqrt{L}} \sum_q c^{\dagger}_k c^{{\phantom{\dagger}}}_{k+q} (a_q^\dagger + a_{-q}) \;,
        \label{Eq:Holstein}
 \end{align}
 where $L$ is the linear size of the system, $c^{\dagger}_k$ and $c_k^{{\phantom{\dagger}}}$ are creation and annihilation operators for electrons, $\epsilon(k)$ is the one-particle electron energy, $a_q^{\dagger}$ and $a_q^{{\phantom{\dagger}}}$ are the creation and annihilation operators of the  phonons, $\omega_0$ is the frequency of the Einstein phonons, $M$ is the oscillator mass,
 %(will be fixed to 1 along this text)
 and $g$ is the strength of the electron-phonon coupling. After linearization around the Fermi energy and introducing  left ($L_k^{{\phantom{\dagger}}}$, $L_k^{\dagger}$) and right movers ($R_k^{{\phantom{\dagger}}}$, $R_k^{\dagger}$), the kinetic energy term becomes
\begin{equation}
	\sum_{k} c^{\dagger}_k c_k \epsilon(k) \rightarrow \sum_k v_f k (R_k^{\dagger} R_k^{{\phantom{\dagger}}} - L_k^{\dagger} L_k^{{\phantom{\dagger}}}).
\end{equation}  
Further  after bosonization \cite{delft1998} with boson operators $b_q^{{\phantom{\dagger}}}$, $b_q^{\dagger}$ this becomes 
\begin{equation}
	\sum_q v_f |q| b_q^{\dagger} b_q^{{\phantom{\dagger}}} \; .
\end{equation}
Neglecting backward scattering one gets -- now also including the phonons:
\begin{align}
	H_{LL} &= \sum_{q=-\infty}^{\infty} v_f |q| b_q^{\dagger} b_q + \omega_0\sum^{\infty}_{q=-\infty} a^{\dagger}_q a_q  \nonumber \\
	&+ \frac{g}{\sqrt{4 \pi \omega_0 M }}  \sum_{q=-{\rm cutoff}}^{q_{\rm cutoff}} \sqrt{|q|} (b_{-q}^{\dagger}+ b_q)(a_q^{\dagger} + a_{-q}) \; .
\label{eq:hamiltonian}
\end{align}
Here, there is a new parameter $q_{\rm cutoff}$ since the Luttinger description is only valid in the low energy sector, so the high-momentum states are secluded from the electron-phonon interaction \cite{varga1964}.   
In the following, we will investigate the Hamiltonian defined in Eq.~(\ref{eq:hamiltonian}), the Holstein model 
        (\ref{Eq:Holstein}) merely serves as a motivation. The main differences are the (infinite) linear electron dispersion  of Hamiltonian (\ref{eq:hamiltonian}), and the lack of the backscattering term. 
In the following, we set $\omega_0=M=v_F=1$.

\section{Canonical transformation}
\label{sec:cantrans}

In this Section, we diagonalize the model (\ref{eq:hamiltonian}) using a Bogoliubov transformation, calculate the pair correlation functions and from this in turn  reconstruct the density matrix of the electrons in terms of the bosonized operators.
 In the literature of entanglement measures of free bosonic systems, canonical impulse and coordinate operators are used more often than the creation and annihilation operators $b^\dagger$, $b$ of the bosons, mainly due to historical reasons. To fit to this major part of the literature we define cosine and sine (or even and odd)  canonical modes for the lattice. 
Specifically, the cosine modes for the lattice read
\begin{eqnarray}
	\hat{Q}_{C,q}=&\frac{1}{2\sqrt{\omega_0}}(a^{\dagger}_q + a_{q} + a^{\dagger}_{-q} + a_{-q})\; ,\\
	\hat{P}_{C_q}=& \frac{i \sqrt{\omega_0}}{2}(a^{\dagger}_q - a_{q} + a^{\dagger}_{-q} - a_{-q}) \; ,
\end{eqnarray}
and the sine modes 
\begin{eqnarray}
	\hat{Q}_{S,q}=&\frac{1}{2\sqrt{\omega_0}}(a^{\dagger}_q + a_{q} - a^{\dagger}_{-q} - a_{-q})\; ,\\
	\hat{P}_{S_q}=&\frac{i \sqrt{\omega_0}}{2}(a^{\dagger}_q - a_{q} - a^{\dagger}_{-q} + a_{-q}) \; ,
\end{eqnarray}
representing cosine and since real space movements, respectively.

We also introduce cosine/sine modes in the electronic subsystem, however we would like to mention that there is no simple connection between these operators, and the electron movements (only to a modulation of the electron density). 
For the electrons, the cosine modes  read
\begin{eqnarray}
	\hat{q}_{C,q}=&\frac{1}{2\sqrt{v_F |q|}}(b^{\dagger}_q + b_{q} + b^{\dagger}_{-q} + b_{-q}) \; ,\\
\hat{p}_{C_q}=& \frac{i \sqrt{ v_F |q|}}{2}(b^{\dagger}_q - b_{q} + b^{\dagger}_{-q} - b_{-q}) \; , \label{eq:el_cosine_modes}
\end{eqnarray}
and the sine modes
\begin{eqnarray}
	\hat{q}_{S,q}=&\frac{1}{2\sqrt{v_F |q|}}(b^{\dagger}_q + b_{q} - b^{\dagger}_{-q} - b_{-q}) \; ,\\
	\hat{p}_{S_q}=&\frac{ i \sqrt{v_F |q|}}{2}(b^{\dagger}_q - b_{q} - b^{\dagger}_{-q} + b_{-q}) \; . \label{eq:el_sine_modes}
\end{eqnarray}

The Hamiltonian then takes the following form  
\begin{align}
	H=&  \frac{1}{2} \sum_{q>0} \left[ \hat{p}_{C,q}, \hat{P}_{C,q} \right] \left[\begin{array}{cc}
		1 & 0 \\
		0 & 1
	\end{array} \right] \left[  \begin{array}{c}
		\hat{p}_{C,q} \\ \hat{P}_{C,q}
	\end{array}\right] \nonumber \\
	&+  \left[ \hat{q}_{C,q}, \hat{Q}_{C,q} \right] \left[\begin{array}{cc}
	 (v_F |q|)^2 &  g q\sqrt{\frac{2  v_F}{ \pi M}} \\
		g q \sqrt{\frac{2  v_F}{\pi M}} & \omega_0^2
	\end{array} \right] \left[  \begin{array}{c}
		\hat{q}_{C,q} \\ \hat{Q}_{C,q}
	\end{array}\right] \nonumber \\
	& + \left[ \hat{p}_{S,q}, \hat{P}_{S,q} \right] \left[\begin{array}{cc}
		1 & \frac{g}{\omega_0} \sqrt{\frac{2}{ \pi M v_F }}  \\
		\frac{g}{\omega_0} \sqrt{\frac{2}{ \pi M v_F }}  &1 
	\end{array} \right] \left[  \begin{array}{c}
		\hat{p}_{S,q} \\ \hat{P}_{S,q}
	\end{array}\right] \nonumber \\
	& +  \left[ \hat{q}_{S,q}, \hat{Q}_{S,q} \right] \left[\begin{array}{cc}
		(v_F |q|)^2 & 0 \\
		0 &  \omega_0^2
	\end{array} \right] \left[  \begin{array}{c}
		\hat{q}_{S,q} \\ \hat{Q}_{S,q}
	\end{array}\right] 
\end{align}
The sine (cosine) mode of the lattice couples only to the sine (cosine) mode of the electrons. Next, we introduce new canonical variables $\hat{p}_{1,q}, \hat{p}_{2,q}, \hat{p}_{3,q}, \hat{p}_{4,q}$, $\hat{q}_{1,q}, \hat{q}_{2,q}, \hat{q}_{3,q}, \hat{q}_{4,q}$ to diagonalize the Hamiltonian
\begin{equation}
	\left[ \begin{array}{c}
		\hat{p}_{1,q}\\
		\hat{p}_{2,q}
	\end{array} \right] = 
	\left[ \begin{array}{cc} A_q & -B_q \\
		B_q & A_q \end{array} \right] 
	\left[ \begin{array}{c}
		\hat{p}_{c,q} \\ \hat{P}_{c,q}
	\end{array} \right] \; ,
\end{equation}	

\begin{equation}
	\left[ \begin{array}{c}
		\hat{q}_{1,q}\\
		\hat{q}_{2,q}
	\end{array} \right] = 
	\left[ \begin{array}{cc} A_q & -B_q \\
		B_q & A_q \end{array} \right] 
	\left[ \begin{array}{c}
		\hat{q}_{c,q} \\ \hat{Q}_{c,q} 
	\end{array} \right]\; .
\end{equation}	

Here, the $A_q$ and $B_q$ numbers (defined in Eqs.~(\ref{Eq:Ak}),(\ref{Eq:Bk}) below) are the components of the eigenvectors of the momentum matrix of the Hamiltonian. The sine modes can be diagonalized by a slightly modified unitary transformation:
\begin{equation}
	\left[ \begin{array}{c}
		\hat{p}_{3,k}\\
		\hat{p}_{4,k}
	\end{array} \right] = 
	\left[ \begin{array}{cc} A_k \frac{\omega_{3,k}}{v_F k} & -B_k \frac{\omega_{3,k}}{\omega_0} \\
		B_q \frac{\omega_{4,k}}{v_F k}& A_q \frac{\omega_{4,k}}{\omega_0} \end{array} \right] 
	\left[ \begin{array}{c}
		\hat{p}_{s,q} \\ \hat{P}_{s,q}
	\end{array} \right] \; ,
	\label{eq:sinus_trans_p}
\end{equation}	

\begin{equation}
	\left[ \begin{array}{c}
		\hat{q}_{3,q}\\
		\hat{q}_{4,q}
	\end{array} \right] =
	\left[ \begin{array}{cc} A_q \frac{v_F q}{\omega_{3,k}} & -B_q \frac{\omega_0}{\omega_{3,q}} \\
		B_q \frac{v_F q}{\omega_{4,q}} & A_q \frac{\omega_0}{\omega_{4,q}} \end{array} \right] 
	\left[ \begin{array}{c}
		\hat{q}_{s,q} \\ \hat{Q}_{s,q}
	\end{array} \right] \; .
	\label{eq:sinus_trans_q}
\end{equation}	
The thus diagonalized Hamiltonian takes the following simple form of four uncoupled harmonic oscillators:
\begin{equation}
	H=\sum_{q>0} \sum_{i=1}^4 (\frac{1}{2} \hat{p}^2_{i,q} + \frac{1}{2} \omega^2_{i,q} \hat{q}^2_{i,q}) \; .
	\label{eq:ham_diag}
\end{equation}
Here,  $A_q$, $B_q$,  and $\omega_{1,q} \dots \omega_{4,q}$  are:
\begin{align}
	\omega^2_{1,q}=\omega^2_{3,k}&= \frac{(v_F q)^2 \!+\! \omega_0^2}{2} + \sqrt{\frac{((v_F q)^2 \!-\! \omega_0^2)^2}{4} \!+\!\frac{ v_F g^2 q^2}{\pi M}} \; , \\
	\omega^2_{2,k}=\omega^2_{4,q}&=  \frac{(v_F q)^2 \!+\! \omega_0^2}{2} - \sqrt{\frac{((v_F q)^2 \!-\! \omega_0^2)^2}{4} \!+\! \frac{ v_F g^2 q^2}{\pi M}}   \; ;
\end{align}
\begin{align}
	B_q&=-\frac{g q}{ \sqrt{N_{q}} }  \sqrt{\frac{ v_F}{\pi M}}\label{Eq:Bk} \; ,\\	
	A_q&= \frac{1}{\sqrt{N_{q}}} \left( (v_F |q|)^2-\omega^2_1 \right) \label{Eq:Ak} ,\\
	N_{q}&=  q^2\frac{g^2 q^2 v_F}{\pi M} +  \left( (v_F |q|)^2-\omega^2_1 \right)^2 .
\end{align}

We can further rewrite the diagonalized Hamiltonian Eq.~(\ref{eq:ham_diag})
again in terms of  creation and annihilation operators 
\begin{align}
	b_{i,q}&=\sqrt{\frac{\omega_{i,q}}{2}} \left( \hat{q}_{i,q} + \frac{i}{\omega_{i,q}} \hat{p}_{i,q} \right) \; ,\\
	b^{\dagger}_{i,q}&=\sqrt{\frac{\omega_{i,q}}{2}} \left( \hat{q}_{i,q} - \frac{i}{\omega_{i,q}} \hat{p}_{i,q} \right) \;.
\end{align}
With these operators the Hamiltonian becomes
\begin{equation}
  H=\sum_{i,q>0} \omega_{i,q} (b^{\dagger}_{i,q} b_{i,q}^{{\phantom{\dagger}}} +1/2 ) \; ;
\end{equation}
and the excited states can be simply labeled  by the occupation numbers $n_{i,q}$ for the bosonic degrees of freedom:
\begin{equation}
	| \{  n_{i,q} \} \rangle = \prod\limits_{\stackrel{\scriptsize q=0\ldots\pi}{\scriptsize i=1\ldots4}} \; (b^{\dagger}_{i,q})^{n_{i,q}} | 0 \rangle \; .
\end{equation} 
The stability criterion of the system in our notations is $\pi \omega^2_0 v_F M  > g^2$ \cite{wentzel1950,bardeen1951,varga1964}.

 \section{Reduced density matrix}
 \label{sec:density_matrix}
 Next, we calculate the reduced density matrix $S$ of the electrons from the pair correlation functions. Our quadratic Hamiltonian has been diagonalized by a canonical transformation. As a consequence, the Wick theorem holds in any subsystem, and the expectation value of any operator string can be calculated using the pair correlation functions. On the other hand, this means that if we find a Gaussian operator (exponential of a quadratic operator), which reproduces the pair correlation functions when used as a density matrix, this Gaussian operator must  give the correct results for all operators, so it is the real density matrix of the subsystem.  More details can be found in \cite{peschel2012}. 
 Hence the strategy is to calculate all pair correlation functions (there are lot of zeros), and to find an appropriate  Gaussian operator.
   
 To obtain $S$, let us thus calculate every correlation function between the bosonized operators.
  The sine and cosine modes of the electron system are coupled to the corresponding phonon modes, but not to each other. Hence, the non-zero  correlation functions of the electron subsystem are only $\langle q_{C,q}^2 \rangle$, $\langle p_{C,q}^2 \rangle$, $\langle p_{C,q} q_{C,q} \rangle$ and  $\langle q_{S,q}^2 \rangle$, $\langle p_{S,q}^2 \rangle$, $\langle p_{S,q} q_{S,q} \rangle $. The coordinate-momentum correlation functions are constant $\langle p_{C,q} q_{C,q} \rangle=\langle p_{S,q} q_{S,q} \rangle=i$, which is a consequence of the commutator relations.
 
 These correlation functions relate to the correlation functions of the  eigenmodes $i=1\ldots 4$ as follows (any correlation function between different eigenmodes is zero): 
 \begin{align}
 	\langle \hat{q}_{s,q}^2 \rangle &= \frac{\omega^2_{1,q}}{(v_F q)^2} A^2_q \langle \hat{q}^2_{3,q}\rangle + \frac{\omega^2_{2,q}}{(v_F q)^2} B^2_{q} \langle \hat{q}^2_{4,q}\rangle \; ,\\
   \langle \hat{p}_{s,q}^2 \rangle &= \frac{(v_F q)^2}{\omega^2_{1,q}} A^2_q \langle \hat{p}^2_{3,k}\rangle + \frac{(v_F q)^2}{\omega^2_{2,q}} B^2_{q} \langle \hat{p}^2_{4,q} \rangle \; ,\\
   \langle \hat{q}_{c,q}^2 \rangle &= A^2_q \langle \hat{q}^2_{1,q}\rangle +  B^2_{q}  \langle \hat{q}^2_{2,k} \rangle \; ,\\
   \langle \hat{p}_{c,q}^2 \rangle &= A^2_q \langle \hat{p}^2_{1,q}\rangle +  B^2_{q}  \langle \hat{p}^2_{2,q}  \rangle \; .
   \label{eq:corr_func}
 \end{align}
 These are in turn directly related to the
 the expectation values  $\langle b^{\dagger}_{i,q} b_{i,q}^{\phantom{\dagger}}\rangle$,  the bosonic occupation numbers, via 
\begin{align} 
	\langle \hat{q}^2_{i,q}\rangle&= (2 \langle  b^{\dagger}_{i,k} b_{i,q}^{\phantom{\dagger}} \rangle +1)/(2\omega_{i,k}) \\
	\langle \hat{p}^2_{i,q}\rangle&=\omega_{i,k} (2\langle  b^{\dagger}_{i,q} b_{i,q}^{\phantom{\dagger}} \rangle+1)/2  
\end{align}
For the ground state, excited states and thermal ensemble these expectation values are
 \begin{equation}
 	\langle  b^{\dagger}_{i,q} b_{i,q}^{{\phantom{\dagger}}} \rangle = \left\{ \begin{array}{ccc}
 		0 & & \textrm{in the ground state} \\
 		n_{i,q} & & \textrm{in excited state } | \{ n_{i,q}\} \rangle \\
 		\frac{1}{\exp(\beta \omega_{i,q})-1} & & \textrm{at temperature } T=1/\beta 
 	\end{array}\right.
 \end{equation}
Now we listed all pair correlation functions. Since there is  no (non-trivial) correlation between the sine and the cosine modes of the lattice, the density matrix has to be a product of a sine-mode and cosine mode part as well as a product between different momenta $q>0$, i.e.:
%\begin{strip}
\begin{align}
	\rho &= \Pi_{q>0}^{q_{\rm cutoff}} (1 - e^{\beta_{c,q}}) e^{-\sum_q \beta_{c,q} B^{\dagger}_{c,q} B_{c,q}} \nonumber\\
	& \times \Pi_{q>0}^{q_{\rm cutoff}}(1 - e^{\beta_{s,q}}) e^{-\sum_q \beta_{s,q} B^{\dagger}_{s,q} B_{s,q}} \nonumber\\
	& \times \rho_{q>q_{\rm cutoff}}
	\label{eq:density_op} \; .
\end{align}
%\end{strip}
Here  $B^{\dagger}_{s,q}$,  $B_{s,q}$, $B^{\dagger}_{c,q}$,  $B_{c,q}$ are bosonic creation and annihilation operators, which can be defined in the following way
\begin{align}
	B^{\dagger}_{s,q}&=\sqrt{\frac{\alpha_{s,q}}{2}} (\hat{q}_{s,q}+ \frac{i}{\alpha_{s,q}} \hat{p}_{s,q})  \\
	B_{s,q}&=\sqrt{\frac{\alpha_{s,q}}{2}} (q_{s,q}- \frac{i}{\alpha_{s,q}} \hat{p}_{s,q}) \\
	B^{\dagger}_{c,q}&=\sqrt{\frac{\alpha_{c,q}}{2}} (\hat{q}_{c,q}+ \frac{i}{\alpha_{c,q}} \hat{p}_{c,q} ) \\
	B_{c,q}&=\sqrt{\frac{\alpha_{c,q}}{2}} (\hat{q}_{c,q}- \frac{i}{\alpha_{c,q}} \hat{p}_{c,q} ) \label{eq:canonical} 
\end{align}
where $\alpha_{C,q}$, $\alpha_{S,q}$, $\beta_{C,q}$, $\beta_{S,q}$ are unknown parameters. These coefficients are the only remaining "free" parameters of the density matrix, because of the high symmetries (lot of zero correlations).
One  has to choose these parameters in way, that the correct pair correlation  functions as calculated above are restored. 
%We know, that the Wick theorem holds for the operators of the electron subspace, and expectation value of any operator string can be expressed by the pair correlation functions listed in  Eq.~(\ref{eq:corr_func}).  With  the form of the density operator in the Eq.~(\ref{eq:density_op}), it is guaranteed that the Wick theorem holds: One only has to set the $\alpha_{s,q}$, $\alpha_{c,q}$, $\beta_{s,q}$, $\beta_{c,q}$ parameters in a way, that the density matrix (\ref{eq:density_op}) \kh{[does what?]}.
Now we calculated all pair correlation functions using  the form Eq.~(\ref{eq:canonical})  as the function of the unknown parameters (not shown here explicitly).   

\begin{figure}
	\includegraphics[width=\linewidth]{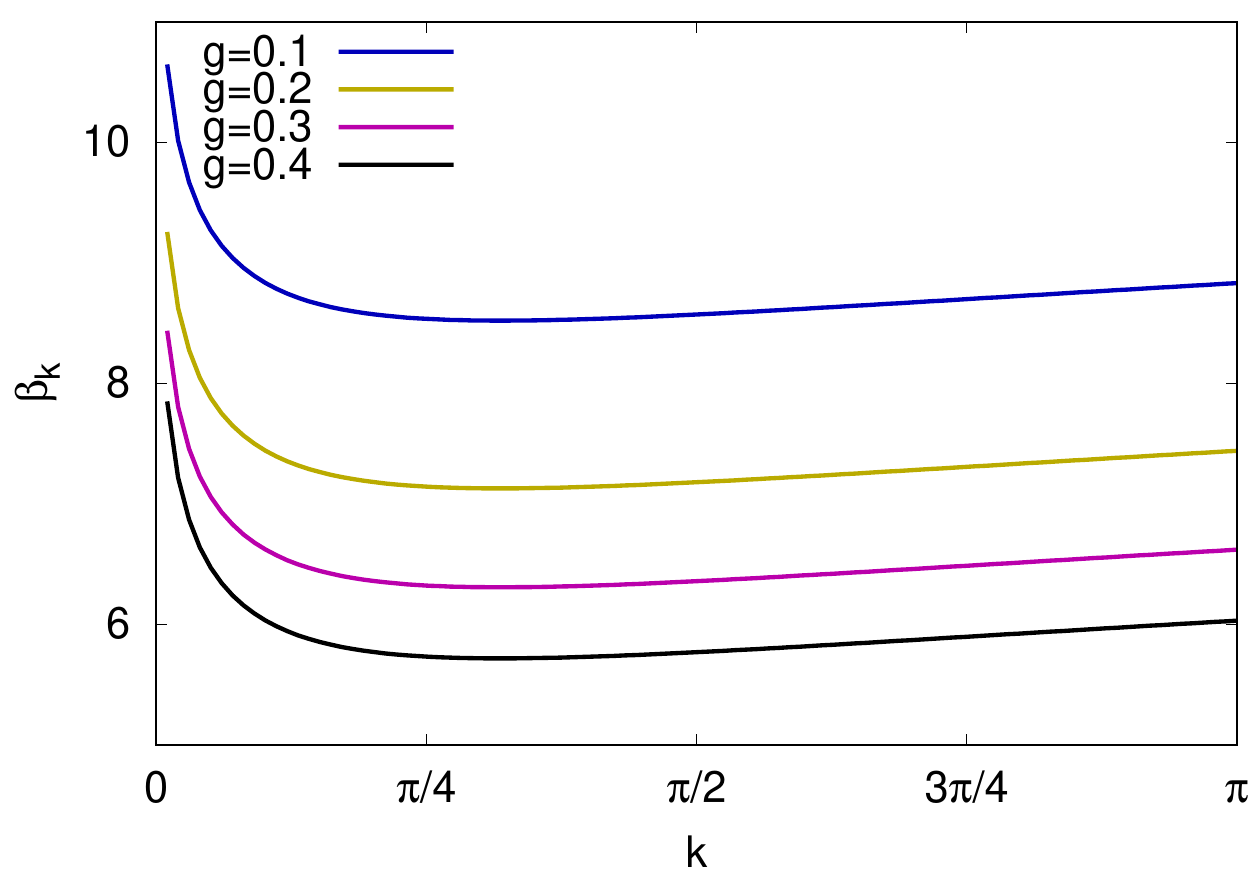}
	\includegraphics[width=\linewidth]{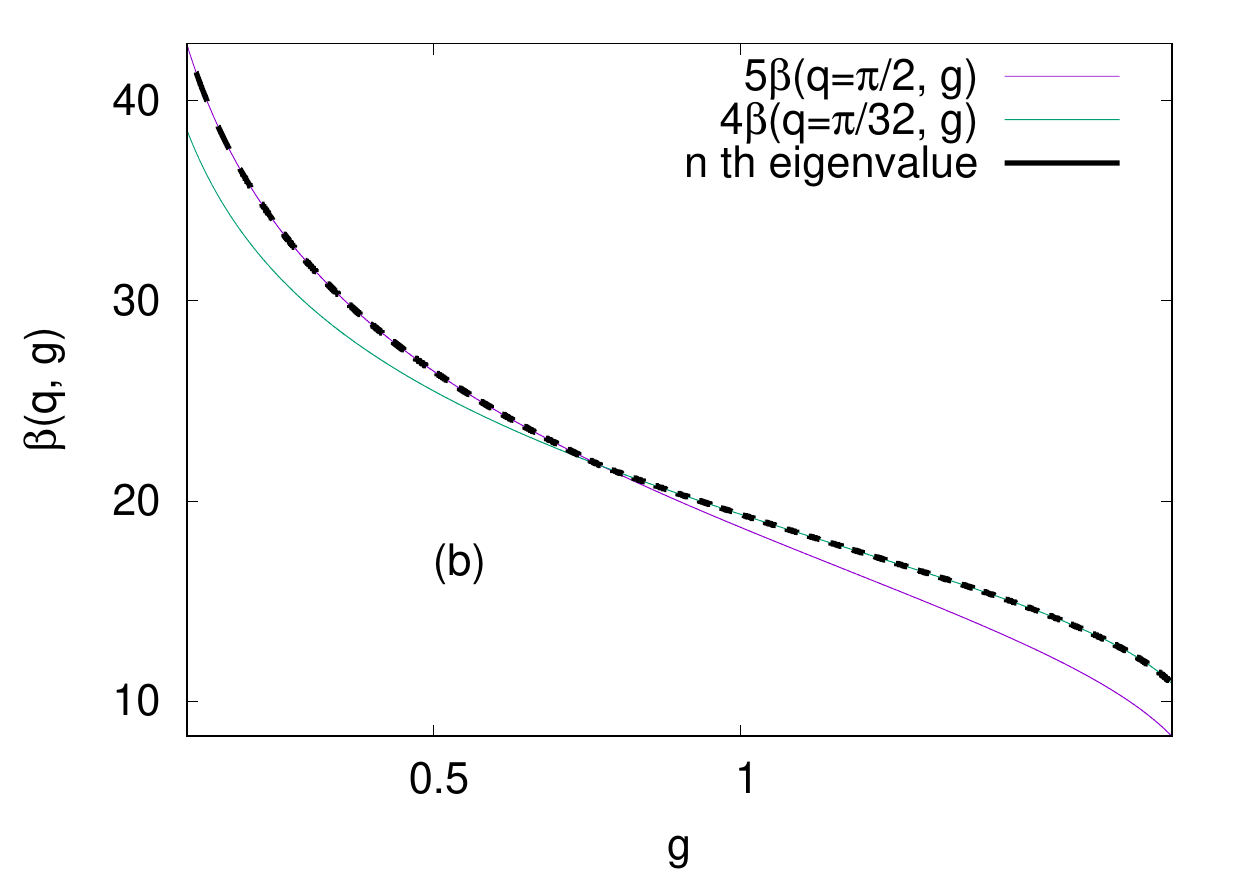}
	\caption{(a) One-particle spectrum $\beta_{C,k}=\beta_{S,k}=\beta_k$ of the entanglement Hamiltonian for various coupling strengths $g$ at zero temperature.  (b) Crossing of different eigenvalues  of the entanglement Hamiltonian. Here and in the following $\omega_0=M=v_F=1$.}
	\label{fig:ent_spectrum_T0}
\end{figure}

We first find that the root of $ {\langle \hat{p}^2_{S/C,q}\rangle }/{\langle \hat{q}^2_{S/C,q} \rangle}$ is $\alpha_{C/S,q}$.  Then we realized, that the product of these (${\langle \hat{p}^2_{S/C,q}\rangle } {\langle \hat{q}^2_{C/S,q} \rangle}$) is related to the occupation numbers.  With these findings all parameters can be determined as follows:

\begin{align}
	\alpha_{C/S,k}&=\sqrt{\frac{\langle \hat{p}^2_{S/C,k}\rangle }{\langle \hat{q}^2_{S/C,k} \rangle}} \; ,\\
	\beta_{C/S,k}&=\ln\left(1+\frac{1}{\sqrt{\langle \hat{p}^2_{S/C,k}\rangle \langle \hat{q}^2_{S/C,k} \rangle } -1/2}\right) \; . \label{eq:beta}
\end{align}
These equations together with similar ones for the reduced density matrix of the phonons presented in Appendix \ref{sec:DMphonons}  form the main result of this paper.

Generally, we have $\alpha_{C,k} \neq \alpha_{S,k}$, and $\beta_{S,k}\neq\beta_{C,k}$, but at zero temperature  $\beta_{S,k}=\beta_{C,k}$. 
The bosonic $B_{s,q}$ and $B_{c,q}$ operators are related to the original bosonized $b_q$ operators and the ladder operators of the Harmonic oscillator through a Bogoliubov transformation.

From  Eq.~(\ref{eq:canonical})   it further follows that $\beta_{s,q}$, $\beta_{c,q}$ are the one-particle eigenvalues of the entanglement Hamiltonian $H_{\rm ent}=-\log \rho$, which is a bosonic free particle Hamiltonian.
This one-particle spectrum of the entanglement Hamiltonian for various coupling strengths is shown in Fig.~\ref{fig:ent_spectrum_T0} (a).
Please note that
the full entanglement spectrum is not simply the set of the one-particle eigenvalues of the entanglement Hamiltonian. Even in our case of a free bosonic  entanglement Hamiltonian, the spectrum of$H_{\rm ent}$ still  includes all multiples of the one-particle eigenvalues (and all combination of different multiples).

Although the one-particle eigenvalues (as a function of the coupling strength) do not cross each other, there can hence still be level crossings in the full entanglement spectrum. An illustrative example is shown in Fig.~\ref{fig:ent_spectrum_T0} (b).
If one arranges the entanglement eigenvalues by magnitude, the crossing simply 
implies a break in the derivative of the nth entanglement eigenvalue [but the eigenvalue itself remains a continuous function of the coupling, see Fig.~\ref{fig:ent_spectrum_T0} (b).]

In \cite{vladimir2020} it has been found by exact diagonalization of a Peierls-type Hamiltonian, that the entanglement spectrum is a nonanalytical function of the coupling strength, the $n$th  entanglement eigenvalue is a continuous function of the coupling strength, but its derivative is not continuous.
Our integrable model gives a simple explanation at least for some similar singularities. Due to the finite bandwidth of the Holstein Hamiltonian, singularities of a different origin may appear. Indeed not all singularities presented in \cite{vladimir2020} seems to be consistent with level crossing. On the other hand, the results of \cite{vladimir2020} result from exact diagonalization; and despite the usage of state-of-art methods, only a very small part of the entanglement spectrum can be explored with exact diagonalization. 
Hence, we do think that the crossing presented here exist also in finite bandwidth systems, but additional singularities may originate from th band-edge.

%In the zero temperature limit, the one-particle energies of the entanglement spectrum can be written in a relative simple form using the following expectation values,

%\begin{align}
%\langle q_{s,k}^2 \rangle &= \frac{\omega_{1,k}}{(v_F k)^2} C^2_k  + %\frac{\omega_{2,k}}{(v_F k)^2} D^2_{k}  \\
%\langle p_{s,k}^2 \rangle &= \frac{(v_F k)^2}{\omega_{1,k}} C^2_k  + \frac{(v_F %k)^2}{\omega_{2,k}} D^2_{k} \\
%\langle q_{c,k}^2 \rangle &= A^2_k \langle q^2_{1,k}\rangle \frac{1}{\omega_{1,k}}+  %B^2_{k} \frac{1}{\omega_{2,k}} \langle q^2_{2,k} \rangle \\
%\langle p_{c,k}^2 \rangle &= A^2_k \omega_{1,k} +  B^2_{k}  \omega_{2,k}
%\label{eq:corr_func_T_null}
%\end{align}

\begin{figure}
	\includegraphics[width=\linewidth]{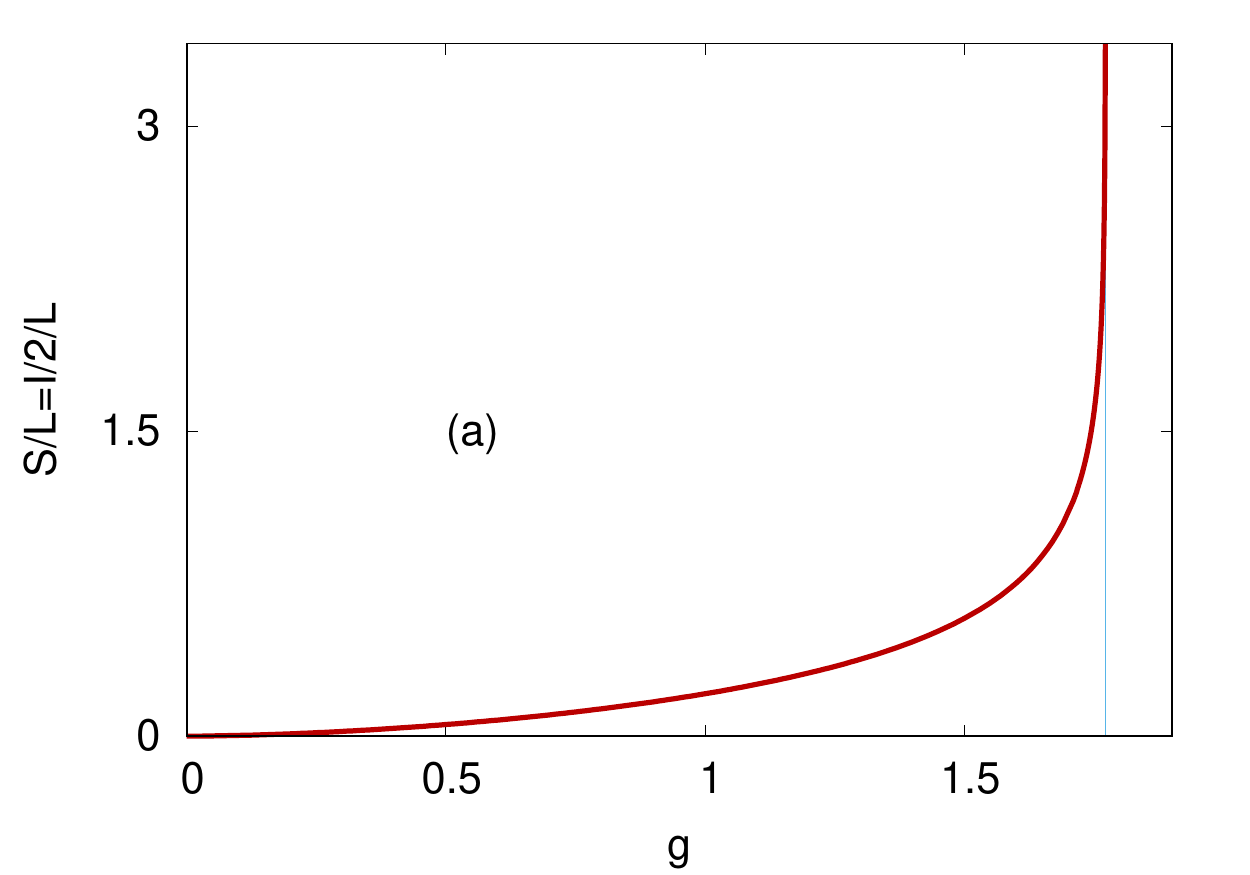}
	\includegraphics[width=\linewidth]{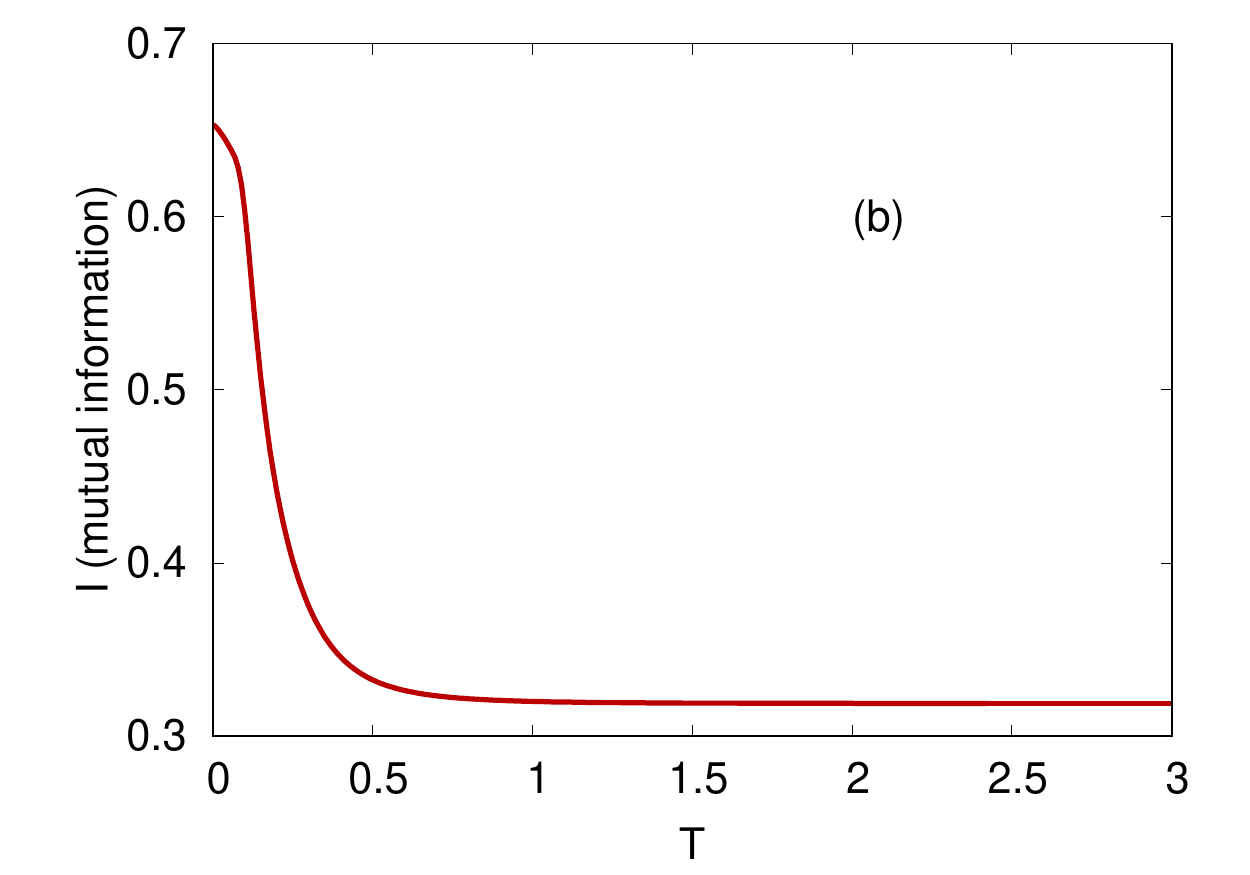}
	\caption{Upper panel: Entanglement entropy $S$ for the electronic subsystem (with a bipartition into  electrons and phonons) as the function of the coupling strength $g$ for the ground state. Lower panel: Mutual information $I$ as the function of  temperature at interaction $g=0.1$.
        \label{Fig:ent_measures}}
\end{figure}

\section{Entanglement measures at zero and finite temperature}
\label{sec:ent_mes}
In this, section we present result for the entanglement entropy
\begin{equation}
  S\equiv S_e=- {\rm Tr} \rho \log \rho
  \label{eq:S}
\end{equation}
for the ground and thermally excited state with the reduced density operator $\rho$ for   the electronic subsystem given by Eq.~(\ref{eq:density_op}), and for
the mutual information
\begin{equation}
  I=S_e + S_{\rm ph}-S_{tot} \; ,
\end{equation}
where  $S_{\rm ph}$ and $S_{tot}$ are in analogy to Eq.~(\ref{eq:S}) the entanglement entropies of the phonon  (see Appendix \ref{sec:DMphonons}) and total system, respectively. The results for the present model are shown in Fig.~\ref{Fig:ent_measures} (see Appendix \ref{sec:vNentropy} for calculational details).  The entanglement entropy $S$ grows with the electron-phonon coupling  strength $g$ and diverges at the Wentzel-Bardeen singularity. At zero temperature [panel (a)], the mutual information is $I=2S$ since for our bipartition $S_e= S_{\rm ph}$ and $S_{tot}=0$. However, at finite temperatures $T>0$ [panel (b)], neither of the latter two equations hold and there are deviations. Specifically, the mutual information decreases with increasing temperature,
and after a local minimum around $T\approx 0.6$  reaches a plateau value for $T\to \infty$.

\begin{figure}
	\includegraphics[width=\linewidth]{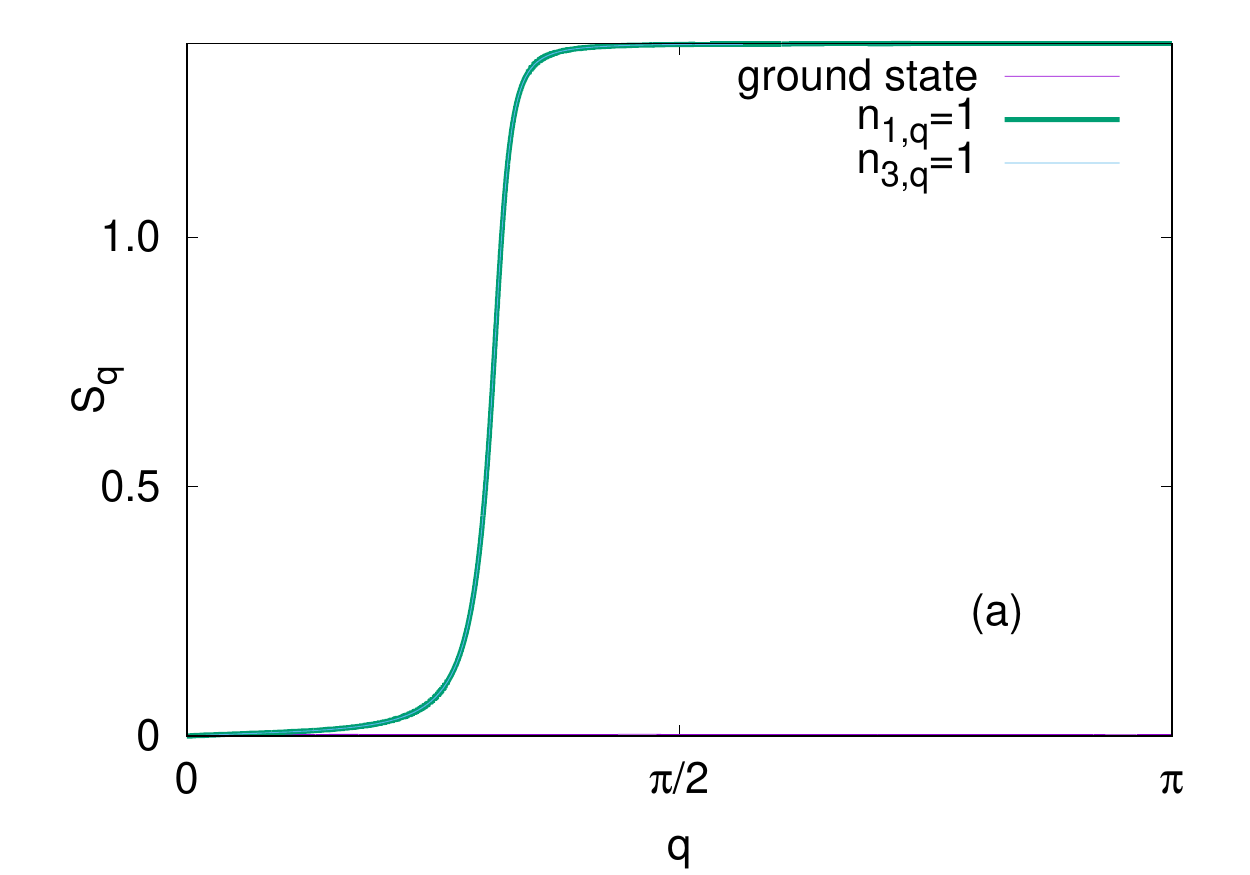}
	\includegraphics[width=\linewidth]{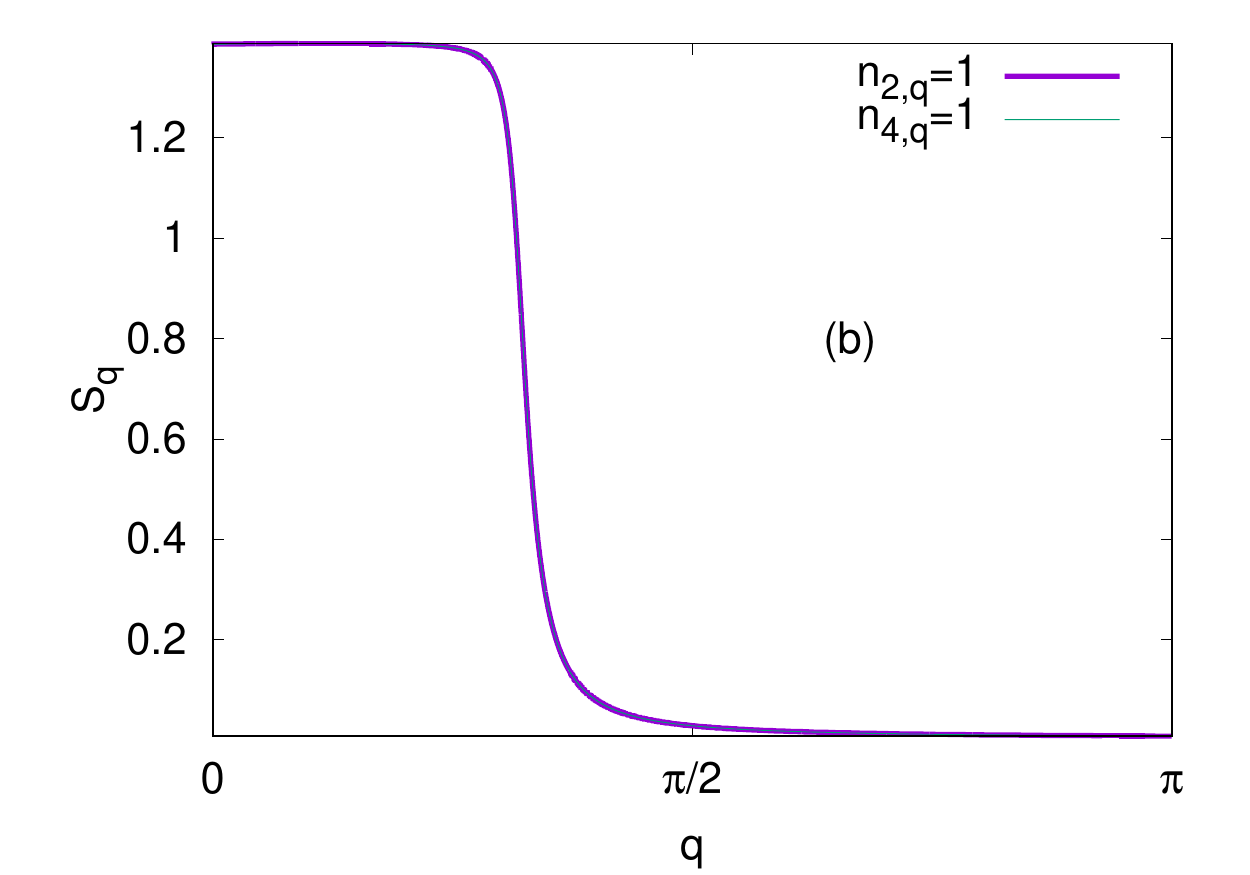}
	\includegraphics[width=\linewidth]{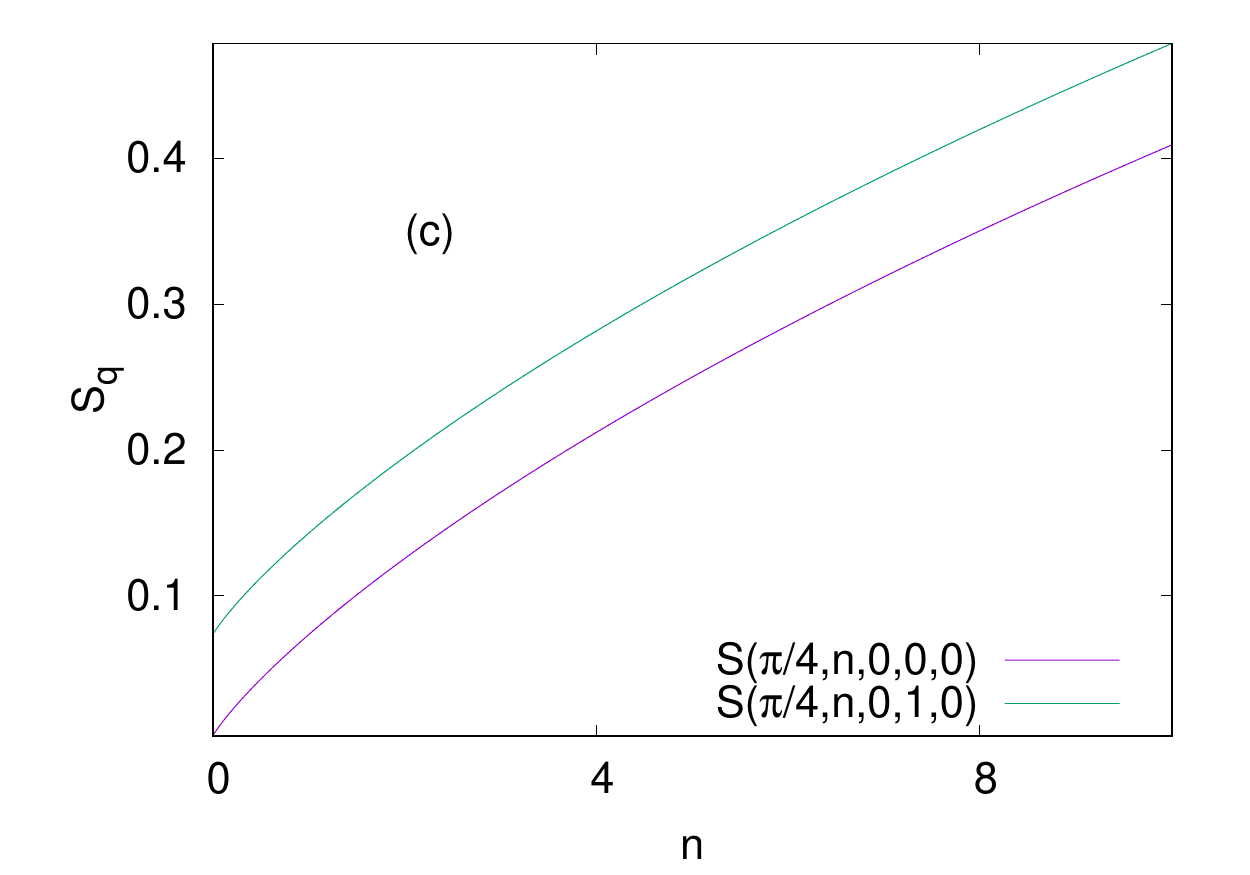}
	\caption{(a) Entanglement  content of the mode with momentum $q$ in the ground state, and in the excited state with $n_{1,q}=1$ ($n_{3,q}=1$) and all other occupation numbers zero. Here and for all subpanels $g=0.1$. (b) Entanglement  content of the mode with momentum $q$  in the excited states where $n_{2,q}=1$ ( $n_{4,q}=1$) and all other occupation numbers are zero. 
	  (c) Entanglement content of the excited state with $n_{1,\pi/4}\neq 0$ and all other occupation number zero as the function of the occupation number $n=n_{1,\pi/4}$ (purple line). The green line shows the entanglement content of the excited state where  $n_{1,\pi/4}\neq0$ and, in addition, $n_{3,\pi/4}=1$. 
          The sum of the entanglement contents of the two states with only one non-zero occupation number (blue line) differs from the state where both
          are excited simultaneously (green line).\label{fig:e_states} }
\end{figure}
\begin{figure}
	\includegraphics[width=\linewidth]{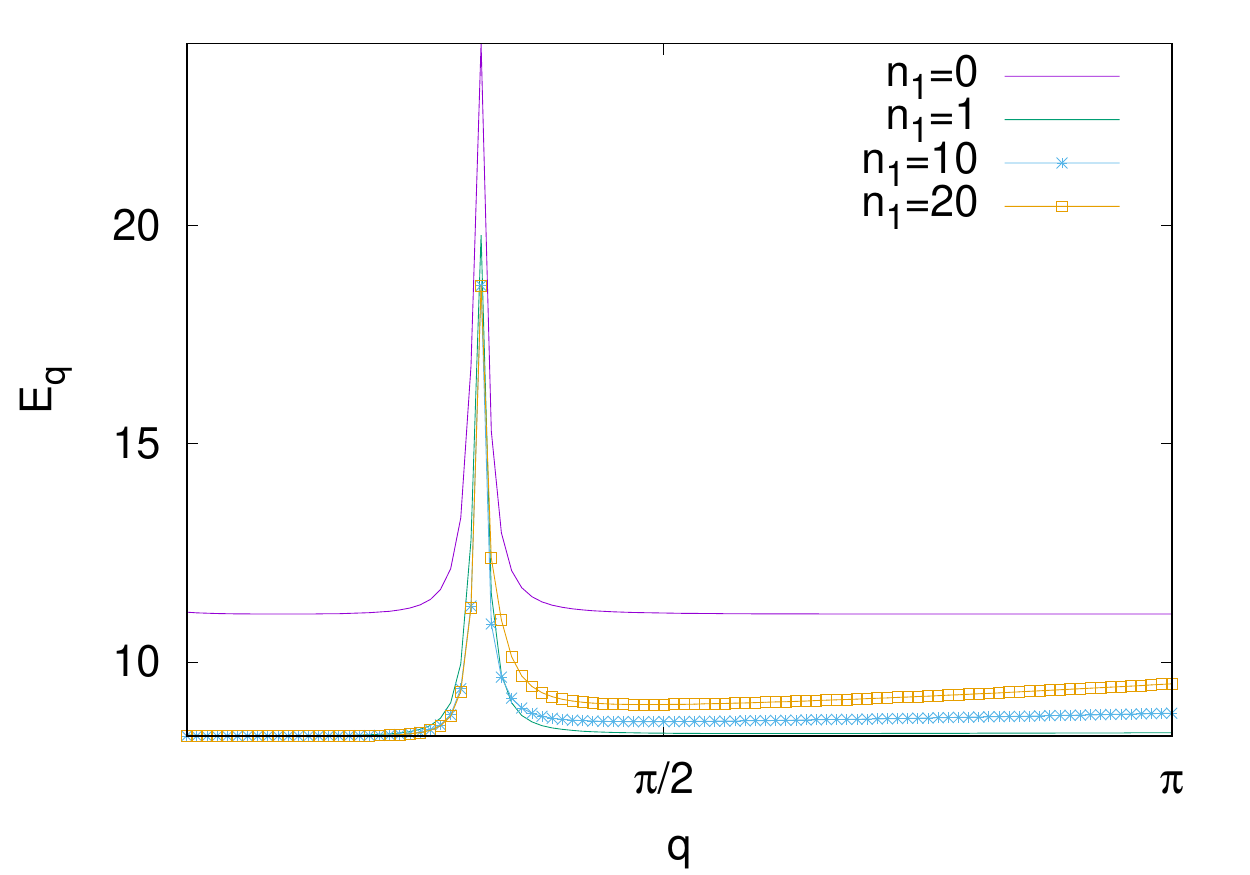}
	\caption{Logarithmic entanglement negativity content of modes with a given wave number at coupling $g=0.1$; $v_F=M=\omega_0=1$. \label{Fig:negativity}  }
\end{figure}
%\begin{eqnarray}
%	\beta_{s,q} &= \\
%	\beta_{c,q} &=
%\end{eqnarray}
\section{Entanglement content of excited states}
\label{sec:ent_exc}
In this section, we investigate the electron-phonon entanglement content of excited states $| \{ n_{i,k} \} \rangle$.
The entanglement entropy can be written as  (see Appendix \ref{sec:vNentropy})
\begin{equation}
	S=\sum_{q>0} s(\langle \hat{p}^2_{C,q}  \rangle \langle \hat{q}^2_{C,q}  \rangle) + 
	s(\langle \hat{p}^2_{S,q}  \rangle \langle \hat{q}^2_{S,q}  \rangle)
	\label{eq:ex_exc}
\end{equation}
with $s(x)=(\sqrt{x}+1/2)\ln(\sqrt{x}+1/2)-(\sqrt{x}-1/2)\ln(\sqrt{x}-1/2)$. It is clear from Eq.~(\ref{eq:ex_exc}), that the entanglement content of the different momenta is additive $S=\sum_q S_q$.  This agrees with Ref.~\onlinecite{szecsenyi2018} and confirms Eq.~(\ref{Eq:kdep}), and further  shows that a generalization to non-spatial bipartitions is possible. 
However, the entanglement content of  the excitations  $b^{\dagger}_{i,q}$ at the same momentum is not linearly additive for  $i=1 \dots 4$. If it was additive, the green and blue lines in  Fig.~\ref{fig:e_states} (c) would coincide. The entanglement is also not proportional  to the occupation number as observed in Ref.~\onlinecite{pizorn2012}: clearly    the red line in  Fig.~\ref{fig:e_states} (c) is not linear.
The entanglement content of the 1st and 3rd and the 2nd and 4th kind of excitations are equal, see panels (a) and (b) of Fig. \ref{fig:e_states}.

The entanglement negativity can be written as the following sum over the momentum
\begin{equation}
	{\cal E} = - 2 \sum_{q=0,\pm}^{q_{cutoff}} \ln \textnormal{min}(1, \sqrt{\Lambda_{C,q, \pm} })+\ln \textnormal{min}(1, \sqrt{\Lambda_{S,q, \pm} }) ,
\end{equation}
the derivation and the $\Lambda_{C/S,k, \pm} $ values are given in  Appendix \ref{sec:neg}. In Fig.~\ref{Fig:negativity} the terms corresponding to a given momentum $k$ are shown for different occupation numbers $n_1$. Quite surprisingly, the  contribution of  a given mode to the entanglement negativity varies in a non-monotonic way with the occupation number. For small occupations the negativity become smaller then the ground state negativity, before starting to increase again for high occupation  numbers. This is a rather different behavior than the entanglement entropy. The entanglement entropy always grows wit excitations, one can define a positive entanglement content for each one-particle excitation. As we have seen, the negativity can decrease with one-particle excitations. If one defines a "negativity content of excitations", this content could be positive or negative.     

\section{Conclusion and outlook}
\label{sec:conc}
We have derived an analytical formula for the the entanglement entropy and
the entanglement spectrum of a Luttinger liquid coupled to an Einstein phonon.
The entanglement spectrum is 
$\sum_{q>0} \beta_{C,q} n_{C,q} + \sum_{q>0} \beta_{S,k} n_{S,q}$, where $n_{C,q}=0,\, 1,\, 2, \dots$ and $n_{S,q}=0,\, 1, \, 2, \dots$ are the occupation numbers and $\beta_{C/S,q}$ from Eq.~(\ref{eq:beta}) the eigenvalues of the bosonic modes of the entanglement Hamiltonian. The entanglement spectrum and thus the entanglement entropy is additive in momentum $q>0$. For each $q$ the Luttinger liquid coupled to Einstein phonons has four bosonic eigenmodes $\beta_{i,q}$, $i=1\ldots 4$,
because  $q$ and $-q$ couple into  independent sine ($S$) and cosine ($C$) linear combinations \cite{dora2017}. For each of these in turn, electrons and phonons couple into two bosonic eigenmodes. In terms of the occupation of these four eigenmodes  $\beta_{i,q}$ at fixed $q$, the entanglement entropy is not additive.
In other words, while the momentum  additivity conjectured
in Eq.~(\ref{Eq:kdep}) holds in our model,
there is no such additivity for the  quasi-particle excitations at each momentum. As in exact diagonalization for a Peierls-type Hamiltonian \cite{vladimir2020}, we find that the entanglement spectrum is a nonanalytical function of the coupling strength, caused by a level crossing of the eigenvalues.

In our calculations, we did not include the electron spin. If included, there will be spin modes and charge modes, and only the charge modes couple to the phonon system. Another simple way of generalization is to include the Coulomb interaction between the  electrons, but the main effect of these in a Luttinger liquid is the renormalization of the Fermi velocity.

Our results are predominately of fundamental, theoretical interest.
However, for prospective applications let us mention that similar one-dimensional electron-phonon models  describe the low energy behavior of carbon nanotubes  \cite{suzuura2002, rosati2015, martino2003, benyamini2014}  and the surface states of thin topological insulator wires \cite{egger2010,dorn2020}.
There are also three-dimensional systems such as  Li$_{0.9}$Mo$_6$O$_{17}$ \cite{giamarchi2012}  where the electron system is quasi one-dimensional, and can be modeled as several parallel chain, each described by the Luttinger theory.
% More interesting direction of future work may be application to real physical systems, carbon nanotubes, topological insulator nanowires are 1D  systems with Luttinger-like electron behaviour and electron phonon coupling, so
For these systems, our results provide a first, general point of understanding, which need to be further detailed to actually describe these materials. Another non-trivial  generalization for higher dimension is the inclusion the phonons to the coupled-wire description of topological systems \cite{meng2020}.

\begin{acknowledgments} G. R.  thanks Carsten Timm and Zoltan Zimbor\'as and Ferenc Igl\'oi useful discussions.
  This work was supported in part  by the National Quantum Information Laboratory of Hungary, and by the National Research, Development and Innovation Office NKFIH under Grant No. K128989.  and the Special Research Programme the (SFB)
  F~86  “Correlated Quantum Materials and Solid-State Quantum Systems” of the Austrian  Science Funds (FWF).
\end{acknowledgments}

\appendix{
\section{Calculation of the entanglement entropy and the mutual information}
\label{sec:vNentropy}
The von Neumann entropy of the reduced density matrix is the entanglement entropy in the case of pure states (e.g.\ at zero temperature). It is also used in the definition of the mutual information in the case of general (for example thermal) states. The calculation of this quantity  has been published in \cite{peschel2012,audenaert2002,roosz2020}, but to be self-contained we summarize it here in a nutshell.

The investigated subsystem contains $N$ bosonic modes with $\hat{q}_1 \dots \hat{q}_N$ and $\hat{p}_1 \dots \hat{p}_N$ canonical conjugated position-momentum operators  with $[q_i, p_j] = i \delta_{i,j}$.  
To calculate the entropy one first evaluates all pair-correlation functions  $Q_{ij}=\langle \hat{q}_i \hat{q}_j \rangle$ and $P_{i,j}= \langle \hat{p}_i \hat{p}_j \rangle$ and $\text{Re} \langle \hat{q}_1 \hat{p}_l \rangle$.
Next, one forms the following correlation matrix of the correlation functions:
\begin{equation}
	M=
	\left(\begin{array}{cccccc}
		\langle \hat{q}_1 \hat{q}_1 \rangle  & \dots & \langle \hat{q}_1 \hat{q}_l \rangle & \text{Re} \langle \hat{q}_1 \hat{p}_1 \rangle & \dots & \text{Re} \langle \hat{q}_1 \hat{p}_l \rangle \\
		\vdots & & \vdots  & \vdots & & \vdots \\
		\langle \hat{q}_l \hat{q}_1 \rangle  & \dots & \langle \hat{q}_l \hat{q}_l \rangle & \text{Re} \langle \hat{q}_l \hat{p}_1 \rangle & \dots & \text{Re} \langle \hat{q}_l \hat{p}_l \rangle \\
		\text{Re} \langle \hat{p}_1 \hat{q}_1 \rangle  & \dots & \text{Re} \langle \hat{p}_1 \hat{q}_l \rangle &  \langle \hat{p}_1 \hat{p}_1 \rangle & \dots &  \langle \hat{p}_1 \hat{p}_l \rangle \\
		\vdots & & \vdots  & \vdots & & \vdots \\
		\text{Re} \langle \hat{p}_l \hat{q}_1 \rangle  & \dots & \text{Re} \langle \hat{p}_l \hat{q}_l \rangle &  \langle \hat{p}_l \hat{p}_1 \rangle & \dots &  \langle \hat{p}_l \hat{p}_l \rangle 
	\end{array} \right).
\label{eq:corr_matrix}
\end{equation}
Now one has to diagonalize the matrix using symplectic transformations and in this way obtain the symplectic spectrum $\Lambda_1 \dots \Lambda_N$. Every eigenvalue of the $M$ matrix is twice degenerated, so $N$ numbers gives the symplectic spectrum of this $2N \times 2N$ matrix. 
Having the eigenvalues of the matrix in hand, one gets the entanglement entropy of the reduced density matrix as

\begin{equation}
	S=\sum_{j=1}^N (\Lambda_j+1/2) \ln(  \Lambda_j +1/2) - (\Lambda_j -1/2) \ln (\Lambda_j -1/2) . 
\end{equation}
  In our model, every coordinate-impulse correlation function is zero $\text{Re} \langle \hat{q}_1 \hat{p}_l \rangle =0$.
  In this case, it can be shown that the square of the symplectic eigenvalues are the eigenvalues of the $PQ$ matrix, where $P$ is matrix containing all momentum-momentum functions (all of $\langle \hat p_{S,k} \hat p_{S,q} \rangle $ and   $\langle \hat p_{C,k} \hat p_{C,q} \rangle $), and $Q$ is a matrix built up from the coordinate correlation functions in a similar manner.
  
  When we consider the sinus-consinus modes defined in   Eq.~(\ref{eq:el_cosine_modes}) and  Eq.~(\ref{eq:el_sine_modes}), the $QP$ matrix further become diagonal, and in the diagonal there are the  $\langle \hat p_{S,k}^2 \rangle$ and the  $\langle \hat p_{C,k}^2 \rangle$ values. This leads to the simple integral formula of Eq.~(\ref{eq:ex_exc}).

   \section{Density matrix of the phonons}
   \label{sec:DMphonons}
With a similar reasoning as we used in the main text to derive the density matrix of the electrons, one can obtain the reduced density matrix of the phonons.

 \begin{align}
 	\rho &= \Pi_{q>0}^{q_{\rm cutoff}} (1 - e^{\beta^{Phonon}_{c,q}}) e^{-\sum_q \beta^{Phonon}_{c,q} D^{\dagger}_{c,q} D_{c,q}} \nonumber\\
 	& \times \Pi_{q>0}^{q_{\rm cutoff}}(1 - e^{\beta^{Phonon}_{s,q}}) e^{-\sum_q \beta^{Phonon}_{s,q} D^{\dagger}_{s,q} D_{s,q}} \nonumber\\
 	& \times \rho_{q>q_{\rm cutoff}}^{\pi}
 	\label{eq:density_op2} \; .
 \end{align}
 %\end{strip}
 Here  $D^{\dagger}_{s,q}$,  $D_{s,q}$, $D^{\dagger}_{c,q}$,  $D_{c,q}$ are bosonic creation and annihilation operators, defined in the following way
 \begin{align}
 	D^{\dagger}_{S,q}&=\sqrt{\frac{\gamma_{S,q}}{2}} (\hat{Q}_{S,q}+ \frac{i}{\gamma_{S,q}} \hat{P}_{S,q}) \; ,  \\
 	D_{S,q}&=\sqrt{\frac{\gamma_{S,q}}{2}} (Q_{S,q}- \frac{i}{\gamma_{S,q}} \hat{P}_{S,q}) \; ,\\
 	D^{\dagger}_{C,q}&=\sqrt{\frac{\gamma_{C,q}}{2}} (\hat{Q}_{C,q}+ \frac{i}{\gamma_{C,q}} \hat{P}_{c,q} )\; ,   \\
 	D_{C,q}&=\sqrt{\frac{\gamma_{C,q}}{2}} (\hat{Q}_{C,q}- \frac{i}{\gamma_{C,q}} \hat{P}_{C,q} )\; ,  \label{eq:canonical2} 
 \end{align}
 where the   parameters $\gamma_{C,q}$, $\gamma_{S,q}$, $\beta_{C,q}$, $\beta_{S,q}$ are obtained in a similar way as in the case of the density matrix of the electrons, yielding  
 \begin{align}
 	\gamma_{C/S,q}&=\sqrt{\frac{\langle \hat{P}^2_{S/C,q}\rangle }{\langle \hat{Q}^2_{S/C,q} \rangle}} \; ,\\
 	\beta^{Phonon}_{C/S,q}&=\ln\left(1+\frac{1}{\sqrt{\langle \hat{P}^2_{S/C,q}\rangle \langle \hat{Q}^2_{S/C,q} \rangle } -1/2}\right) \; . \label{eq:beta2}
 \end{align}
 Here the expectation values of the squares of the bosonic operators are the following
 \begin{align}
 	\langle \hat{Q}^2_{C,q} \rangle = A^2_k \langle \hat{q}^2_{2,q} \rangle + B^2_k \langle \hat{q}^2_{1,q} \rangle \; , \\
 	\langle \hat{P}^2_{C,q} \rangle = A^2_k \langle \hat{p}^2_{2,q} \rangle + B^2_k \langle \hat{p}^2_{1,q} \rangle \; , \\
 	\langle \hat{Q}^2_{S,q} \rangle = A^2_k \frac{\omega^2_{4,q}}{\omega^2_0} \langle \hat{q}^2_{4,q} \rangle + B^2_k \frac{\omega^2_{3,q}}{\omega^2_0} \langle \hat{q}^2_{3,q} \rangle  \; , \\
 	\langle \hat{P}^2_{S,q} \rangle = A^2_k \frac{\omega^2_0}{\omega^2_{4,q}} \langle \hat{p}^2_{4,q} \rangle + B^2_k \frac{\omega^2_0}{\omega^2_{3,q}} \langle \hat{p}^2_{3,q} \rangle   \; .
 \end{align}
 
 \section{Calculation of the entanglement negativity}
 \label{sec:neg}
  To calculate the entanglement negativity between the electrons and the phonons, one can consider the correlation matrix of the whole system, which is similar to  $M$ in Eq.~(\ref{eq:corr_matrix}) but now contains all variables of the phonons and the bosonized 
 fermions. Then one multiples all phonon momenta with $-1$. The thus transformed correlation matrix is the correlation matrix of the partial transpose of the density matrix.

One has to calculate the symplectic eigenvalues of the transformed matrix, let us denote these eigenvalues as $\lambda$.
Then, the logarithmic negativity is calculated as 
\begin{equation}
	{\cal E}=-\sum_{\lambda} \ln \min(1,\lambda)
\end{equation}
In our problem the correlation matrix is block-diagonal, so one can find the symplectic eigenvalues block-by-block, and write the negativity as a sum over momentum.
\begin{equation}
	{\cal E} = - 2 \sum_{q=0,\pm}^{q_{\rm cutoff}} \ln \textnormal{min}(1, \sqrt{\Lambda_{C,q, \pm} })+\ln \textnormal{min}(1, \sqrt{\Lambda_{S,q, \pm} }) ,
\end{equation}
where
\begin{equation}
	\Lambda_{C/S, q, \pm} = \frac{1}{2} \left[a_{C/S,q} \pm  \sqrt{a_{C/S,q}^2 +4 b_{C/S,q} -4 c_{C/S,q}} \right] 
\end{equation}
with
\begin{widetext}
{\small \begin{eqnarray} 
	a_{C/S,q} &=& \langle \hat Q_{C/S,q} \hat Q_{C/S,q} \rangle \langle \hat P_{C/S,q} \hat P_{C/S,q} \rangle
	+ \langle \hat q_{C/S,q} \hat q_{C/S,q} \rangle \langle \hat p_{C/S,q} \hat p_{C/S,q} \rangle + 2 \langle \hat P_{C/S,q} \hat p_{C/S,q} \rangle  \langle \hat Q_{C/S,q} \hat q_{C/S,q} \rangle , \\
	b_{C/S,q} &=& \left( \langle \hat Q_{C/S,q} \hat q_{C/S,q} \rangle \langle \hat P_{C/S,q} \hat P_{C/S,q} \rangle
	- \langle \hat q_{C/S,q} \hat q_{C/S,q} \rangle \langle \hat P_{C/S,q} \hat p_{C/S,q} \rangle  \right) \nonumber\\
	&& \times \left(\langle \hat Q_{C/S,q} \hat q_{C/S,q} \rangle  \langle \hat p_{C/S,q} \hat p_{C/S,q} \rangle
	- \langle \hat P_{C/S,q} \hat p_{C/S,q} \rangle   \langle \hat Q_{C/S,q} \hat Q_{C/S,q} \rangle \right) , \\
	c_{C/S,q} &=& \left(\langle \hat Q_{C/S,q} \hat Q_{C/S,q} \rangle \langle \hat P_{C/S,q} \hat P_{C/S,q} \rangle
	- \langle \hat P_{C/S,q} \hat p_{C/S,q} \rangle  \langle \hat Q_{C/S,q} \hat q_{C/S,q} \rangle \right) \nonumber \\
	&& \times \left(\langle \hat q_{C/S,q} \hat q_{C/S,q} \rangle \langle \hat p_{C/S,q} \hat p_{C/S,q} \rangle -
	\langle \hat P_{C/S,q} \hat p_{C/S,q} \rangle  \langle \hat Q_{C/S,q} \hat  q_{C/S,q} \rangle \right) .
	\label{eq:ent_res}
\end{eqnarray}}
\end{widetext}

\end{document}